\documentclass[conference]{IEEEtran}
\IEEEoverridecommandlockouts
\usepackage{cite}
\usepackage{amsmath,amssymb,amsfonts}
\usepackage{algorithmic}
\usepackage{graphicx}
\usepackage{textcomp}
\usepackage{xcolor}
\usepackage{stfloats}
\usepackage{float} 
\usepackage{hyperref}
\usepackage{subfigure}
\usepackage{framed}
\usepackage{blindtext}
\usepackage{multicol}
\usepackage{mdwlist}
\usepackage{epstopdf}
\usepackage{color}
\usepackage{fancyhdr}
\usepackage{tikz}

\usepackage[ruled,linesnumbered,vlined]{algorithm2e}

\def\BibTeX{{\rm B\kern-.05em{\sc i\kern-.025em b}\kern-.08em
    T\kern-.1667em\lower.7ex\hbox{E}\kern-.125emX}}
    
\newcommand\copyrighttext{%
  \footnotesize \textcopyright 2022 IEEE. Personal use of this material is permitted.
  Permission from IEEE must be obtained for all other uses, in any current or future 
  media, including reprinting/republishing this material for advertising or promotional 
  purposes, creating new collective works, for resale or redistribution to servers or 
  lists, or reuse of any copyrighted component of this work in other works. 
  DOI: }
\newcommand\copyrightnotice{%
\begin{tikzpicture}[remember picture,overlay]
\node[anchor=south,yshift=10pt] at (current page.south) {\fbox{\parbox{\dimexpr\textwidth-\fboxsep-\fboxrule\relax}{\copyrighttext}}};
\end{tikzpicture}%
}

\begin{document}

\title{Conference Paper Title*\\
{\footnotesize \textsuperscript{*}Note: Sub-titles are not captured in Xplore and
should not be used}
\thanks{Identify applicable funding agency here. If none, delete this.}
}

\author{\IEEEauthorblockN{Wenxuan Yu\IEEEauthorrefmark{2}, 
Minghui Xu\IEEEauthorrefmark{2}\IEEEauthorrefmark{1}, 
Dongxiao Yu\IEEEauthorrefmark{2}, 
Xiuzhen Cheng\IEEEauthorrefmark{2}, 
Qin Hu\IEEEauthorrefmark{3}, 
Zehui Xiong\IEEEauthorrefmark{4}}

\IEEEauthorblockA{\IEEEauthorrefmark{2}School of Computer Science and Technology, Shandong University, China}
\IEEEauthorblockA{\IEEEauthorrefmark{3}Department of Computer and Information Science, Indiana University-Purdue University Indianapolis, USA}
\IEEEauthorblockA{\IEEEauthorrefmark{4}Information Systems Technology and Design Pillar, Singapore University of Technology and Design, Singapore}
\thanks{*Corresponding author: Minghui Xu (mhxu@sdu.edu.cn).}
}

\title{zk-PCN: A Privacy-Preserving Payment Channel Network Using zk-SNARKs}

\maketitle  
\copyrightnotice

\begin{abstract}
Payment channel network (PCN) is a layer-two scaling solution that enables fast off-chain transactions but does not involve on-chain transaction settlement. PCNs raise new privacy issues including balance secrecy, relationship anonymity and payment privacy. Moreover, protecting privacy causes low transaction success rates. To address this dilemma, we propose zk-PCN, a privacy-preserving payment channel network using zk-SNARKs. We prevent from exposing true balances by setting up \textit{public balances} instead. Using public balances, zk-PCN can guarantee high transaction success rates and protect PCN privacy with zero-knowledge proofs. Additionally, zk-PCN is compatible with the existing routing algorithms of PCNs. To support such compatibility, we propose zk-IPCN to improve zk-PCN with a novel proof generation (RPG) algorithm. zk-IPCN reduces the overheads of storing channel information and lowers the frequency of generating zero-knowledge proofs. Finally, extensive simulations demonstrate the effectiveness and efficiency of zk-PCN in various settings. 
\end{abstract}


\section{Introduction}

Blockchains are growing in popularity and have played an important role in the financial ecosystem. The increasing number of users escalates the blockchain scalability problem. In recent years, the research community and industry have proposed several promising scaling solutions. Existing solutions can be divided into on-chain scaling (e.g., sharding and DAG-based blockchain) and off-chain scaling (e.g., payment channel, sidechain). 
Payment channel as a novel off-chain scaling solution is to provide transaction processing capability off-chain and settle finality of transactions on-chain. Therefore, payment channels enable high throughput without requiring consensus on each transaction. This method essentially shifts workloads from blockchain to off-chain channels.

A payment channel network (PCN) is a collection of bidirectional payment channels. A sender can pay any recipient through intermediary channels without taking the trouble to open new direct ones. Lightning Network (LN) \cite{poon2016bitcoin} is the largest payment channel network with 18k nodes and 88k channels \footnote{ Data is obtained from \href{https://ln.fiatjaf.com/}{https://ln.fiatjaf.com/} [2022-07-01].}. PCNs improve the privacy of blockchains since off-chain transactions are not transparently recorded on-chain. However, PCNs raise new privacy issues. With the information from off-chain channels, malicious nodes can obtain users' private information such as account balances \cite{herrera2019difficulty}, payment paths \cite{kappos2021empirical}, and social relationship \cite{erdin2021evaluation}. 

Some efforts have been made to address privacy concerns \cite{green2017bolt, tang2020privacy}. However, there are still two issues to be solved.
Firstly, transaction success rates could be negatively impacted by privacy protection methods. Some PCNs (e.g., LN) only publish channel capacity (i.e., the sum of two account balances of a channel) but hides specific balances. However, hiding balances causes the routing discovery protocol less efficient and thus reduces success rates. 
Tang \textit{et al.} \cite{tang2020privacy} let each channel publish noisy balances besides channel capacity. Still, adding noises could make the balances displayed inaccurate, which leads to transaction failures.  It remains an open problem that how to achieve privacy protection and availability when designing a balance model. 

Secondly, PCNs can not guarantee three elementary privacy properties (i.e., balance secrecy, relationship anonymity, and payment privacy) at once. 
These three properties are formally defined in \cite{kappos2021empirical}. Specifically, balance secrecy states that the balances of a channel are private to others. Relationship anonymity requires that each intermediary node knows nothing about other participants on a transaction route but only its predecessor or successor. Payment privacy guarantees that a node cannot reveal the path of a transaction which they do not participate in. LN intends to preserve balance secrecy with hided balances and achieve relationship anonymity using onion routing. However, the balance discovery attack and the payment discovery attack can still disclose balances and reveal payment paths in LN \cite{herrera2019difficulty}. As for noisy balance based methods \cite{tang2020privacy}, they have the risk of exposing balance secrecy since updating noisy balances might cause payment privacy leakage. 

In this paper, we attempt to address the aforementioned shortcomings with a novel privacy-preserving payment channel network. We highlight our contributions as follows:

\begin{itemize}
\item We propose zk-PCN, a privacy-preserving payment channel network using zk-SNARKs. zk-PCN can protect balance secrecy, relationship anonymity and payment privacy at once. Importantly, zk-PCN still ensures high transaction success rates even with a strong privacy guarantee. 
\item zk-PCN can be compatible with routing algorithms to achieve improvements. To this end, we design the zk-IPCN (`I'' for improved). zk-IPCN integrates the routing algorithm with zk-PCN using a novel reactive proof generation (RPG) algorithm. zk-IPCN reduces the storage and computational overheads of zk-PCN.
\item Extensive simulations of zk-PCN demonstrate that zk-PCN outperforms LN and achieves a high successful rate while protecting privacy. The results also indicate the effectiveness of zk-IPCN.

\end{itemize}

The rest of the paper is organized as follows. We present the related works in section~\ref{RW}, and introduce preliminaries in section~\ref{pre}. In section~\ref{model}, we describe the details of zk-PCN and zk-IPCN. Finally, we evaluate our methods in Section~\ref{impletation} and conclude this paper in Section~\ref{conslusion}. 

\section{Related Work}\label{RW}

Some recent works focus on the vulnerabilities of PCNs. Erdin \textit{et al.} \cite{erdin2021evaluation} defined  major properties that should be achieved by private PCNs and analyzed several privacy-preserving PCNs.  
Kappos \textit{et al.} \cite{kappos2021empirical} introduced the balance discovery attack and the payment discovery attack against Lightning Network. Malavolta \textit{et al.} \cite{malavolta2018anonymous} proposed anonymous multi-hop locks (AMHLs) to resist wormhole attacks. For privacy protection, AMHLs protect relationship anonymity using a onion routing. 

Besides, many studies focus on protecting the privacy of PCNs. Bolt \cite{green2017bolt} provided a strong privacy guarantee by establishing payment channels over Zcash, but Bolt is a hub-based payment system that cannot be extended to the scenarios when we need privacy protection for multi-hop payments. TumbleBit \cite{heilman2016tumblebit} also needs a payment hub to protect transaction anonymity, so it is restricted to single-hop payments.
Malavolta \textit{et al.} \cite{malavolta2017concurrency} focused on the privacy and concurrency of PCNs. They presented a multi-hop HTLC to protect balance secrecy but users' balances should be temporarily locked up if transactions fail. 
Tang \textit{et al.} \cite{tang2020privacy} explored the tradeoff between channel efficiency and channel privacy. They adopted noisy balances to protect balance secrecy. 
Twilight \cite{dotan2022twilight} uses the differential privacy to hide the information of a large number of payments and a Trusted Execution Environment (TEE) to ensure that selfish nodes have to follow its protocol.
%
Pickhardt \textit{et al.} \cite{pickhardt2021security} proposed a probabilistic model, which can be used to calculate success rates given a certain transaction. They also show that Lightning Network usually try many times to find a proper payment route due to opaque balances.

Even though existing works cover lots of PCN privacy issues, there is still a lack of a full-fledged scheme which can protect balance secrecy, relationship anonymity and payment privacy at a time while achieving a high transaction success rate. zk-PCN and zk-IPCN can address these shortcomings using verifiable public balances and a reactive proof generation algorithm. 

\section{Preliminaries}\label{pre}

\subsection{Blockchain and Smart Contracts}

A blockchain is a distributed ledger that keeps a consistent and tamper-proof record stored as blocks. Each block contains multiple transactions. Blockchains use consensus algorithms to order transactions consistently among mutually untrusted nodes. The best-known application of blockchain is cryptocurrency, such as Bitcoin \cite{nakamoto2008bitcoin}. 
Inspired by Bitcoin, other blockchain systems have surfaced. Ethereum extends Bitcoin's script to smart contracts written in a (ideally) Turing-complete programming language, Solidity. Smart contracts are computer programs stored on blockchains and are executed by Ethereum Virtual Machine (EVM). Blockchain and smart contract have been applied in lots of scenarios including Internet of Things \cite{xu2022blown}, distributed learning \cite{xu2022spdl}, and cloud computing \cite{xu2022cloudchain}.
Since the computing resources on the blockchain are very limited, Ethereum sets a gas limit to prevent smart contracts from falling into an infinite loop or occupying all resources on the blockchain. When the gas is exhausted, the contract will stop running. Such limits make smart contracts unable to perform complex function calculations.

\subsection{Payment Channel Network}

\begin{figure}[htbp]
\centerline{\includegraphics[scale=0.4]{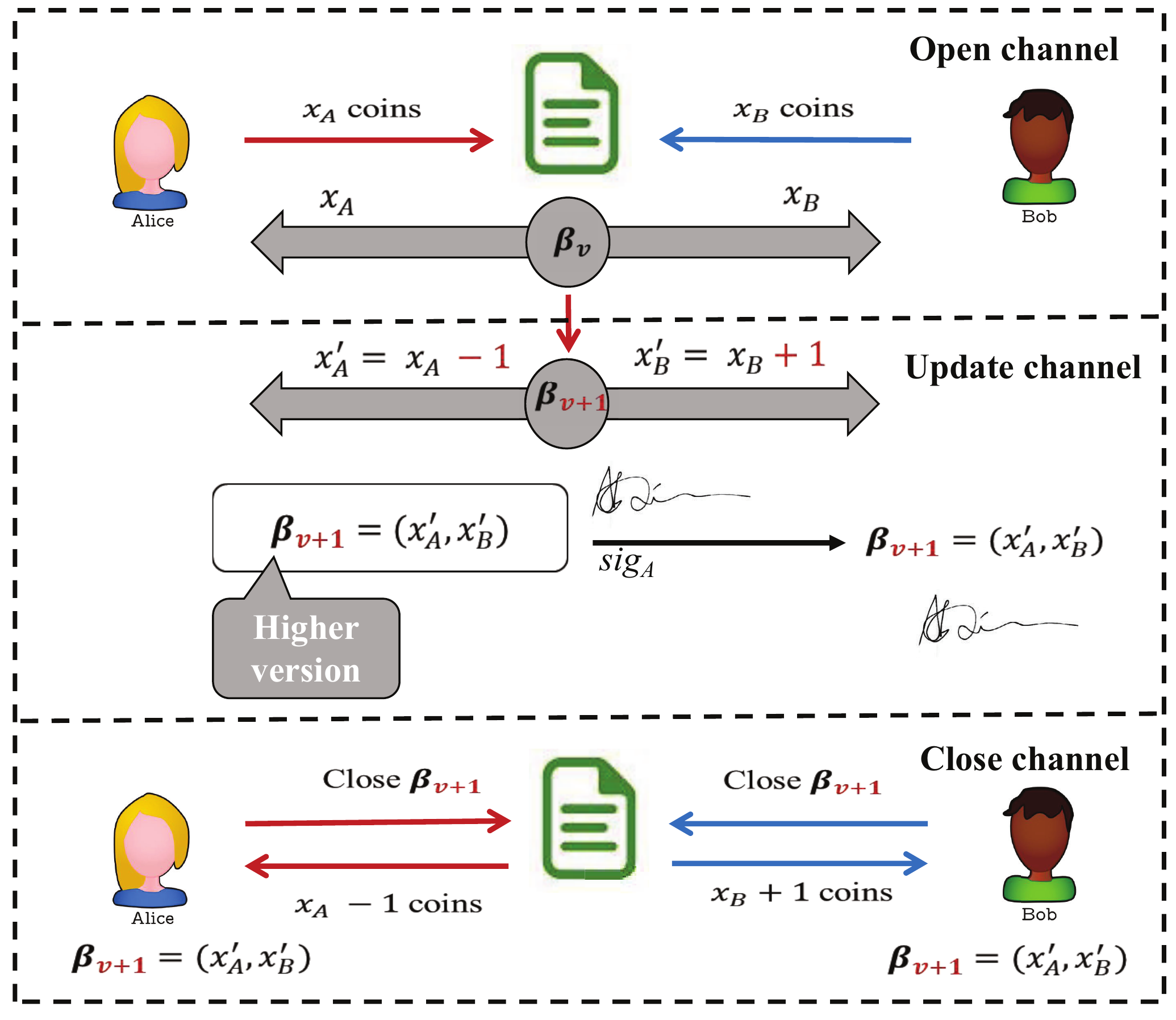}}
\caption{The life cycle of a payment channel.}
\label{payment channel}
\end{figure}

\subsubsection{Payment Channel}

Payment channel is an off-chain scaling technique that improves on-chain performance by moving high-frequency, small-value transactions to off-chain channels. 
As shown in Fig.~\ref{payment channel}, the life cycle of a payment channel has three phases: open channel, update channel and close channel. Here we give an explanation of each phase and leave the illustration of zk-PCN in the next section.

\textbf{Open channel.} Two parties Alice and Bob can collaboratively create a smart contract to open a payment channel $\beta_v$, where $v$ represents the version number and is initially set as $0$. Alice deposits $x_A$ coins and Bob deposits $x_B$ coins into the smart contract. The channel capacity of $\beta_v$ is $x_A+x_B$. 

\textbf{Update channel.} Alice and Bob can send update messages to settle down off-chain payments. For example, when Alice wants to make a payment, she increments the channel version number $v$, and updates the balance allocation to $x_{A}'$ and $x_{B}'$. Then she sends an update message to Bob with her digital signature $\mathsf{sig}_A$. Bob updates his balance allocation if the message received is valid.

\textbf{Close channel.} Each party can unilaterally close the channel and retrieve the channel balance. To this end, Alice can input the latest channel version number with a digital signature to the contract. The contract waits for Bob to determine whether the submitted version number is consistent with himself. If the version numbers are the same, Bob agrees to close the channel. Otherwise, Bob submits his latest channel version number to the contract, which selects the channel state with the highest channel version number. Finally, the contract settles the finality of balance allocation.

\subsubsection{Payment Channel Network}

\begin{figure}[htbp]
\centerline{\includegraphics[scale=0.35]{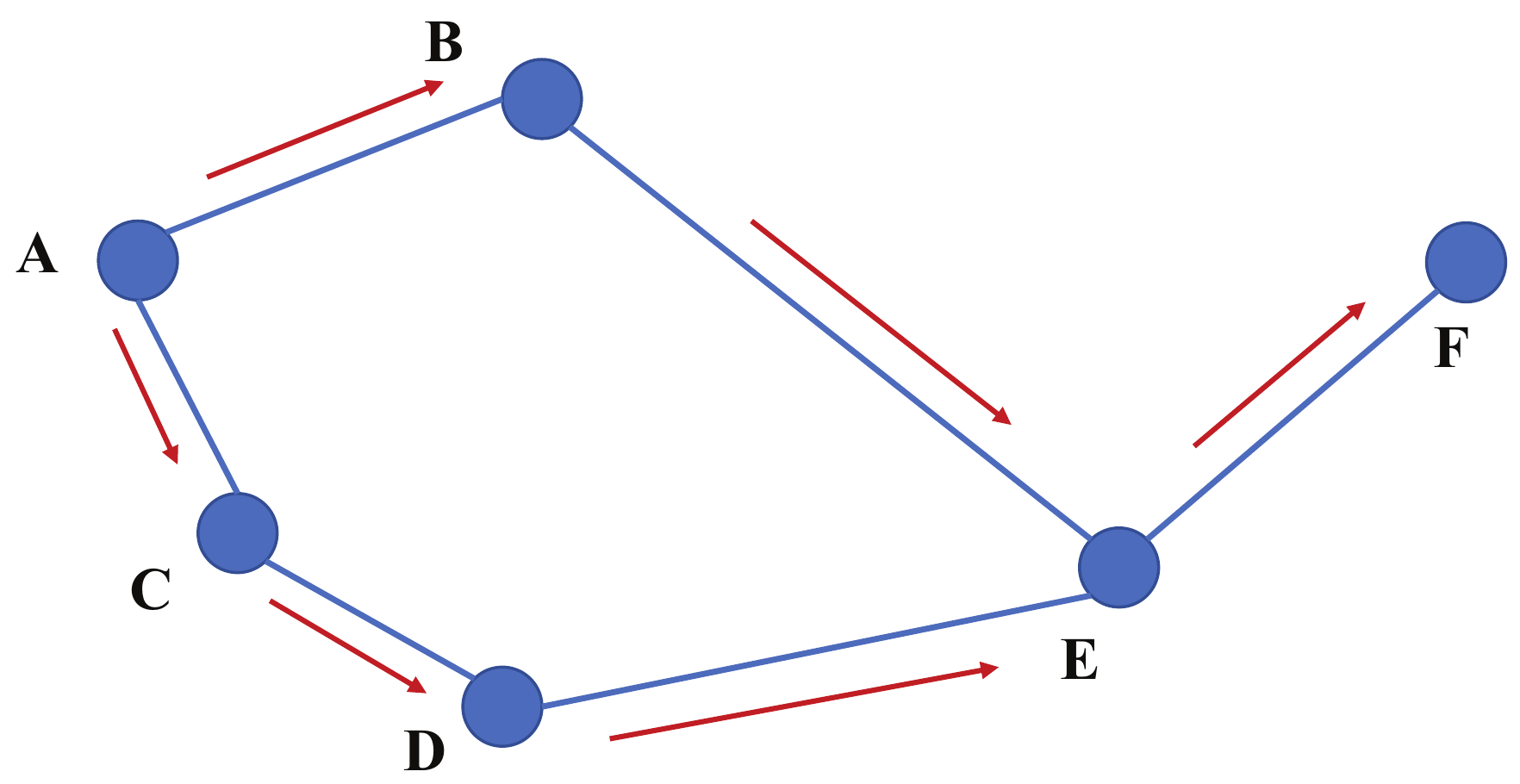}}
\caption{An example of PCN with six nodes.}
\label{payment channel network topology}
\end{figure}

A payment channel network (PCN) is composed of multiple channels. Nodes in a PCN can forward transactions through channels. Intermediate nodes on a multi-hop payment path receive forwarding fee as reward. Fig.~\ref{payment channel network topology} presents an example of payment channel network with six nodes. Edges represent the established channels, and nodes represent the users. For example, A could pay F via path A-B-E-F or A-C-D-E-F. 
The Lightning Network (LN) \cite{poon2016bitcoin} is the largest payment channel network.  LN utilizes Hash Time-Lock Contract (HTLC) to ensure the atomicity of transactions. HTLC adopts hashlocks and timelocks to let the recipient either acknowledge receiving the payment prior to a deadline or forfeit the ability to claim the payment. 

\subsection{Zero-Knowledge Proof and zk-SNARKs}\label{ZKP}

A zero-knowledge proof can prove the correctness of a statement without leaking any private information \cite{goldreich1994definitions}. 
Zero knowledge Succinct Non-interactive Argument of Knowledge (zk-SNARKs) demonstrates its value in privacy protection for blockchain, such as Zerocash \cite{sasson2014zerocash} and Hawk \cite{kosba2016hawk}.  A zk-SNARKs scheme can be represented by a tuple of polynomial-time algorithms $\Pi = (\mathsf{Setup},\mathsf{Prove},\mathsf{Verify})$. zk-SNARKs transform each statement into a circuit $C$, and then work as follows:

\begin{itemize}
    \item $\mathsf{pp}$ $\leftarrow$ $\mathsf{Setup}(\lambda,C)$: The $\mathsf{Setup}$ algorithm takes a security parameter $\lambda$ and a circuit $C$ as inputs, and outputs public parameter $\mathsf{pp}$ consisting of proving key $pk$ and verification key $vk$. 
    \item $\pi$ $\leftarrow$ $\mathsf{Prove}(\mathsf{pp},x,w)$: This algorithm takes public parameter $\mathsf{pp}$, circuit public input $x$ and circuit private input $w$ as inputs, and generates a succinct proof $\pi$ whose size is irrelevant to the circuit size.
    \item $\{0,1\}$ $\leftarrow$ $\mathsf{Verify}(\mathsf{pp},\pi,x)$: This algorithm takes public parameter $\mathsf{pp}$, a succinct proof $\pi$ and circuit public input $x$ as inputs, and outputs the verification result. It outputs 1 if verification is successful, otherwise it outputs 0.
\end{itemize}

Note that each proof is of constant size and efficient to be stored and verified. In addition, zk-SNARKs satisfy succinctness, completeness, soundness, zero-knowledge and non-interactivity \cite{parno2013pinocchio}.

\section{zk-PCN Design}\label{model}

This section presents a detailed description of zk-PCN and zk-IPCN. In Sect.~\ref{CI}, we briefly describe our design goals. Then we illustrate zk-PCN and zk-IPCN in Sect.~\ref{TM} and Sect.~\ref{zk-IPCN}, respectively. 

\subsection{Design Goals}\label{CI}

We aim to achieve the following design goals in developing zk-PCN and zk-IPCN:

\begin{itemize}
    \item \textbf{[G1] Privacy preservation.} Our scheme should satisfy balance secrecy, relationship anonymity and payment privacy properties. Moreover, our privacy protection scheme needs to resist balance discovery attack and payment discovery attack.
    \item \textbf{[G2] High transaction success rates.} Privacy protection might cause channel information unavailable and impacts the transaction success rates. Compared with LN, we require our schemes to have a high transaction success rate while guaranteeing strong privacy protection properties.
    \item \textbf{[G3] Compatibility with routing protocols.}  Routing algorithms are helpful in implementing PCNs \cite{sivaraman2020high, lobmaier2022assessing}. Thus zk-PCN and zk-IPCN should be compatible with existing routing algorithms.
\end{itemize}

\subsection{Privacy-Preserving Payment Channel Network using zk-SNARKs (zk-PCN)}\label{TM}

Here we formally introduce our privacy-preserving payment channel network using zk-SNARKs (zk-PCN). To achieve [G1], zk-PCN adopts a novel balance model inspired by the famous Yao's Millionaires' problem \cite{yao1982protocols}. We set up \textit{public balances} to hide private true balances and thereby protect balance secrecy. The true balance represents the real current account value. Public balance is a fake amount value and made public. Two parties in a channel reveal public balances and a zero-knowledge proof, which proves that each true balance is greater than or equal to the corresponding public balance. 
When a user wants to make a multi-hop payment, the routing algorithm only chooses intermediate channels whose public balances are greater than or equal to the transaction amount. This mechanism helps zk-PCN reach [G2].
Our design prevents malicious nodes from detecting true balances of a channel, and thus resists the balance discovery attack. Moreover, the public balance can be used to confirm the channel availability. 

In zk-PCN, each node maintains a table with each entry representing the initial balance and the public balance of a channel. zk-PCN combines onion routing of Lightning Network to protect relationship anonymity. Therefore the sender of a multi-hop payment can send an anonymous transaction through all intermediate nodes on a path. Additionally, we introduce the details of protecting payment privacy in Section~\ref{update ch}. 

We formally present our zk-PCN protocol in Fig.~\ref{zk-PCN protocol} and describe the details of zk-PCN in the following. 

\begin{figure*}[!htbp]\footnotesize
	
	\begin{framed}
			\centering \textbf{The zk-PCN Protocol Details} \vspace{5pt}
			\begin{basedescript}{\desclabelwidth{30pt}}
				\item[\hspace{7pt}Open Channel:] ~\\ 
				{
				    \setlength{\leftskip}{-16pt}Alice and Bob call $\mathsf{OpenChannel}(pk_A,pk_B,x_A,x_B)$ of smart contract to open a payment channel. \\
				}
				\item[\hspace{7pt}Update Channel:] ~\\
				{
				    \setlength{\leftskip}{-16pt}Alice calculates  
				    $h_i = H(r_i||\mathsf{tran}_i||v_i||\mathsf{sig}_A)$; \\
				    \setlength{\leftskip}{-16pt}Alice sends $\mathsf{update}_{\mathsf{channel}}$ message $\langle\mathsf{sig}_A,v_i,\mathsf{tran}_i,h_i\rangle$ to Bob;\\
					\setlength{\leftskip}{-16pt}Bob updates the balance allocation $(x_A'=x_A-\mathsf{tran}_i,x_B'=x_B+\mathsf{tran}_i)$; \\
					\setlength{\leftskip}{-16pt}Alice gets $\mathsf{pp} = \mathsf{Setup}(1^\lambda,C_{\mathsf{public}})$;\\
					\setlength{\leftskip}{-16pt}Alice generates proof $\pi = \mathsf{Prove}(\mathsf{pp},x,w)$ and broadcasts $\pi$;\\
					\setlength{\leftskip}{-16pt}Alice and Bob ask adjacent nodes to generate and broadcast proofs ;\\
					\setlength{\leftskip}{-16pt}Other nodes verify $\pi$ using $\mathsf{Verify}(\mathsf{pp},x,\pi)$.\\
				}
				\item[\hspace{7pt}Close Channel:]~\\
				{
				\setlength{\leftskip}{-16pt}Alice and Bob call $\mathsf{CloseChannel}(v_n,\mathsf{sig})$ of smart contract to close channel and retrieve the channel balance $(x_A',x_B')$.  \\
				
				}
			\end{basedescript}
			
	\end{framed}\vspace{-10pt}
	\caption{The zk-PCN protocol details. Alice and Bob are the two parties of a payment channel. Alice sends a transaction to Bob. Nodes in the network call our zk-SNARK algorithms $\Delta=(\mathsf{Setup},\mathsf{Prove},\mathsf{Verify})$ to generate and verify proofs. Let $\lambda$ be the security parameter, $C_{\mathsf{public}}$ be the arithmetic circuit in Fig.~\ref{proof algorithm}. Alice (or Bob) is the prover and other nodes in the PCN are verifiers. } \label{zk-PCN protocol}
\end{figure*}


\subsubsection{Open channel}Nodes call the smart contract function $\mathsf{OpenChannel}(pk_A,pk_B,x_A,x_B)$ to open a payment channel. $pk_A$ and $pk_B$ represent Alice's and Bob's public keys recorded on a blockchain. $x_A$ and $x_B$ represent the initial balances of two parties in the channel. Note that only initial balances are true. After channel setup, the true balances of a channel will soon be changed and thus hidden. Therefore, the channel privacy is protected once the channel comes online.

\subsubsection{Update channel}\label{update ch}
\begin{figure}[htbp]
\centering
\centerline{\includegraphics[scale=0.45]{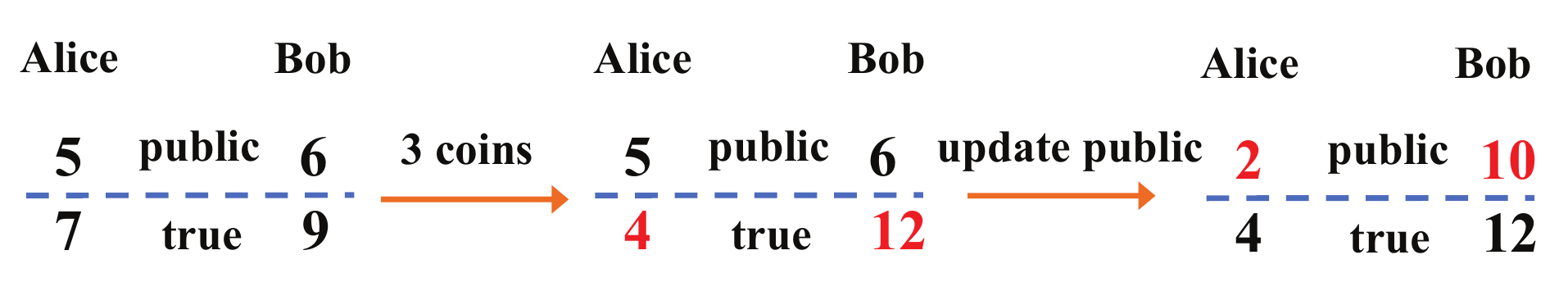}}
\caption{An example of single-round payment channel update.}
\label{update payment channel}
\end{figure}

When a channel updates, the true balances change. A public balance may become higher than the corresponding true balance. Therefore, a channel needs to update the public balance and generate a new zero-knowledge proof which proves that the new public balance is less than or equal to the current true balance. For example, as Fig.~\ref{update payment channel} shows, the true balances of Alice and Bob are 7 and 9, and the public balances are 5 and 6. If Alice sends Bob 3 coins, they need to update balances to satisfy the requirement (public balance $\leq$ true balance).

In detail, if Alice wants to send a transaction, she first generates a hash value $h_i$ used to prove the transaction authenticity. $h_i$ is calculated as follows.
\begin{equation}
h_{i}=H(r_i||\mathsf{tran}_i||v_i||\mathsf{sig}_A)\label{eq},
\end{equation}
where $r_i$ is a random number, $\mathsf{tran}_i$ denotes the transaction amount, $v_i$ represents the latest channel version number, and $\mathsf{sig}_A$ indicates the digital signature of Alice. The subscripts $i$ represents the $i^{th}$ transaction. Then Alice sends $\mathsf{update}_\mathsf{channel}$ message $\langle\mathsf{sig}_A,v_i,\mathsf{tran}_i,h_i\rangle$ to Bob. Alice and Bob each updates their balances accordingly. Subsequently, Alice (or Bob) invokes the circuit $C_{\mathsf{public}}$ to generate a zero-knowledge proof.
Note that Alice and Bob can take turns to generate proofs, which hides the transaction sender and thus protect the flow of transactions.

\begin{figure*}[htbp]
\centerline{\includegraphics[scale=0.50]{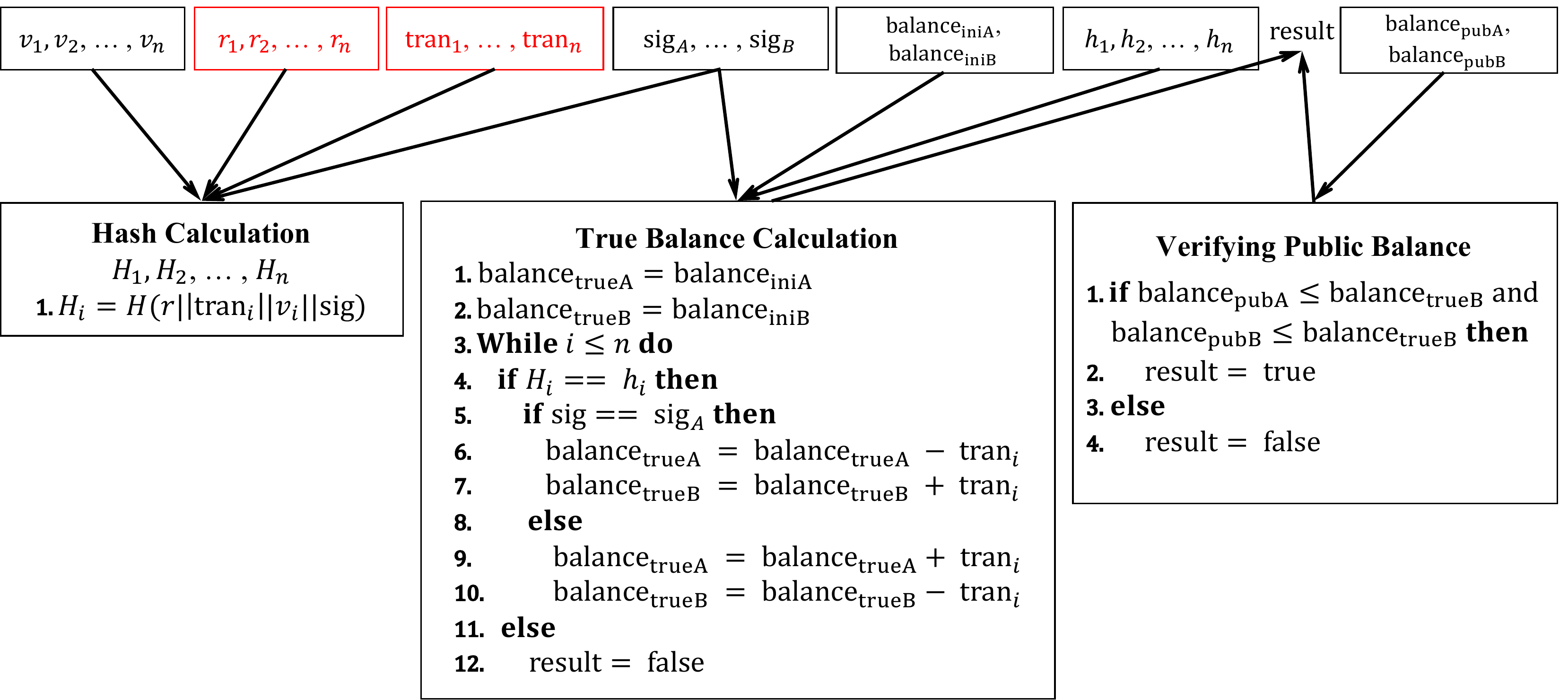}}
\caption{The logic diagram of the circuit used in zk-PCN. Public inputs are in black box, secret witness is in red box.}
\label{proof algorithm}
\end{figure*}

Fig.~\ref{proof algorithm} illustrates the details of the public circuit $C_{\mathsf{public}}$, where black boxes represent public inputs, red boxes represent private inputs, and result is the circuit output.

We input the initial balances ($\mathsf{balance}_{\mathsf{iniA}}$ and $\mathsf{balance}_{\mathsf{iniB}}$) and the hash values of all transactions ($H_1,H_2,...,H_n$) to ensure that the current true balances are calculated from transactions which have not been tampered with. 
Besides, $C_{\mathsf{public}}$ takes ($\mathsf{tran}_1, \cdots, \mathsf{tran}_n$) as private inputs to protect the balance secrecy. 
$C_{\mathsf{public}}$ verifies the hash value of each transaction and calculates the true balance ($\mathsf{balance}_{\mathsf{trueA}}$ and $\mathsf{balance}_{\mathsf{trueB}}$). Finally $C_{\mathsf{public}}$ validates whether public balances ($\mathsf{balance}_{\mathsf{pubA}}$ and $\mathsf{balance}_{\mathsf{pubB}}$) are less than or equal to the true balances. zk-PCN develops three polynomial-time algorithms $\Delta=(\mathsf{Setup},\mathsf{Prove},\mathsf{Verify})$ on the basis of the zk-SNARKs algorithm $\Pi$. 

\begin{itemize}
          \item $\mathsf{pp}$ $\leftarrow$ $\mathsf{Setup}(1^\lambda,C_{\mathsf{public}})$: Given the security parameter $\lambda$ and the circuit $C_{\mathsf{public}}$, $\mathsf{Setup}(1^\lambda,C_{\mathsf{public}})$ generates the public parameter $\mathsf{pp}$. Note that $\mathsf{pp}$ is only bounded to  $C_{\mathsf{public}}$ but independent of the specific inputs.
          \item $\pi$ $\leftarrow$ $\mathcal{P}.\mathsf{Prove}(\mathsf{pp},x,w)$: Prover $\mathcal{P}$ can call this algorithm to generate a proof $\pi$. $x$ and $w$ are the public input and private input of the circuit $C_{\mathsf{public}}$. Nodes can generate proofs with the same $C_{\mathsf{public}}$, $\mathsf{pp}$ and their own inputs.
          \item $\{0,1\}$ $\leftarrow$ $\mathcal{V}.\mathsf{Verify}(\mathsf{pp},x,\pi)$: Verifier $\mathcal{V}$ calls this algorithm to verify the proof $\pi$ with the public input $x$ and the common reference string $\mathsf{pp}$. It outputs 1 if the proof $\pi$ is valid and 0 otherwise.
      \end{itemize}

In payment channel networks, a multi-hop payment path requires updating multiple pairs of public balances. In zk-PCN, a sender decides a multi-hop payment path according to the routing table and public balances. After the sender selects a payment path and successfully forwards the transaction, all channels on the path need to update their public balance accordingly. Each channel broadcasts new public balances and proof $\pi$ to the network. Other nodes can verify them and update entries in their local routing table.

zk-PCN protects payment privacy in two ways. On one hand, a channel can serve multiple different transaction paths at the same time, so it is hard to find a specific path among overlapped ones. On the other hand, a channel which updates its true balances can randomly ask several adjacent channels to update balances. In this way, zk-PCN can prevent other nodes from inferring the transaction path since malicious nodes have no knowledge about which channel serves a specific path. 

\subsubsection{Close channel} Each party of a channel can close the channel unilaterally by calling the smart contract function $\mathsf{CloseChannel}(v_n,\mathsf{sig})$. $v_n$ is the latest channel version number. $\mathsf{sig}$ represents the digital signatures of the latest transaction. If Alice and Bob hold different views of $v_n$, the function will allocate the balance according to the version with a higher number. Closing channel finally returns $x_A'$ and $x_B'$ to Alice and Bob.

\subsection{zk-IPCN}\label{zk-IPCN}

Based on zk-PCN, we propose an improved privacy-preserving payment channel network using zk-SNARKs (zk-IPCN). Recall that the zk-PCN protocol is responsible for channel manipulations. In this section, we aim to show that zk-PCN can be compatible with existing routing algorithms. In zk-IPCN, we introduce a novel reactive proof generation (RPG) algorithm to integrate routing algorithms and zk-PCN while preserving the privacy protection functionality. 

\begin{figure}[!htbp]
\centering
\centerline{\includegraphics[scale=0.4]{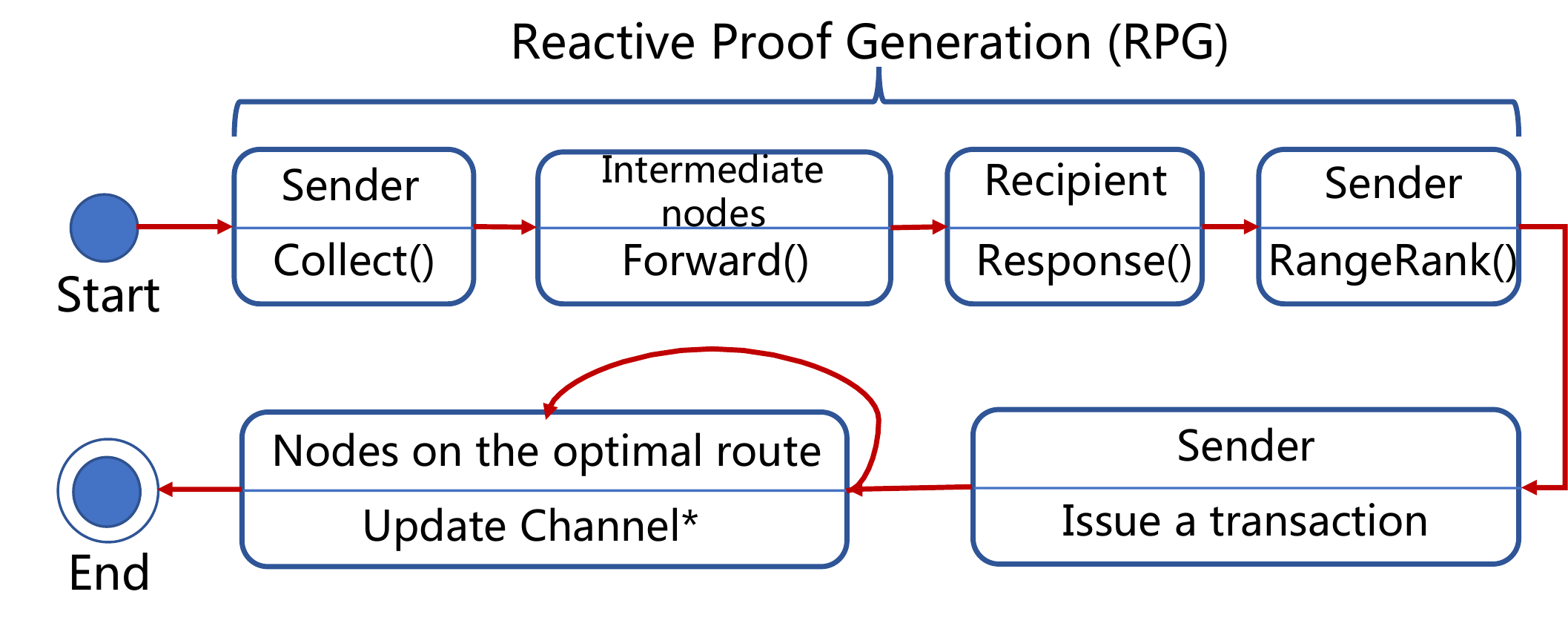}}
\caption{zk-IPCN state diagram.}
\label{state graph}
\end{figure}

Fig~\ref{state graph} shows the entire procedure of zk-IPCN. The RPG algorithm helps a sender find the optimal transaction route with a high success rate and a low forwarding fee. Moreover, zk-IPCN improves the update channel procedure of zk-PCN with a $\mathsf{Update\, Channel}^*$.  In zk-IPCN, $\mathsf{Update\, Channel}^*$ only requires a node to broadcast newly generated proofs to adjacent nodes instead of the entire network. With the RPG algorithm, broadcasting proofs within a single hop still guarantees effective path selections.

With zk-IPCN, nodes only need to store partial channel information, thus preventing from large storage overheads. Moreover, zk-IPCN reduces the number of proofs to be generated and broadcast. The computational overhead and bandwidth usage are thereby reduced. 

The concrete idea of RPG is to classify the information of the payment channel network into static information and dynamic information. Note that a node only needs to store the information related to a local network, namely within $k$-hop \footnote{$k$ is an adjustable constant}.
The static information includes the network topology (presented as a routing table) and the capacity of each channel listed on the table. Dynamic information includes public balances, proofs of public balances and forwarding fees. 
In zk-IPCN, a sender can use an existing routing protocol (e.g., Flare \cite{prihodko2016flare}) to discover all \textit{candidate routes} to a destination, denoted by $P_n=\{p_1, \cdots, p_n\}$.  Then, the node collects all dynamic information on $P_n$ and selects the optimal route. 

There are four functions to be called in the RPG algorithm, namely $\Theta = (\mathsf{Collect},\mathsf{Forward},\mathsf{Response},\mathsf{RangeRank})$. Note that a variable with braces represents a set. For instance, $\pi_i$ represents the $i^{th}$ proof and \{$\pi_i$\} denotes the set $\{\pi_1, \pi_2,\cdots, \pi_i\}$. Besides, we denote a sender, an intermediate node and a recipient by $\mathcal{S}$, $\mathcal{I}$ and $\mathcal{R}$ respectively. $\Theta$ works as follows.

\begin{itemize}
    \item $\mathsf{totalFee}$ $\leftarrow$ $\mathcal{S}.\mathsf{Collect}(\mathsf{tran},P_n)$: The $\mathsf{Collect}$ function takes the transaction amount $\mathsf{tran}$, candidate routes $P_n$ as inputs, and outputs $\mathsf{totalFee}$. $\mathsf{totalFee}$ maps each route to a total forwarding fee being spent on that route.
    The sender transmits $\mathsf{route}_{\mathsf{post}}$ message $\langle\mathsf{tran},\mathcal{S}\rangle$ to collect dynamic information from candidate routes in $P_n$. After receiving enough response messages from recipients (or destinations), the sender stores valid routes and their forwarding fees in $\mathsf{totalFee}$. 
    \item $\{(\pi_j,x_j)\},\mathsf{fee}_{\mathsf{after}}$ $\leftarrow$
    $\mathcal{I}.\mathsf{Forward}(\{(\pi_{j-1},x_{j-1})\},\mathsf{fee})$. Given the previous $j-1$ proofs \{$\pi_{j-1}$\} and public inputs \{$x_{j-1}$\}, a intermediate node calls $\mathsf{Forward}$ function to append its own proof $(\pi_j,x_j)$, and add its charge on $\mathsf{fee}$ to obtain $\mathsf{fee}_{\mathsf{after}}$. Then the intermediate node transfer the outputs $\{(\pi_j,x_j)\}$ and $\mathsf{fee}_{\mathsf{after}}$ to the next stop. 
    \item $\{(\pi_{\hat{j}},x_{\hat{j}})\},\mathsf{fee}$ $\leftarrow$ $\mathcal{R}.\mathsf{Response}(\{(\pi_{\hat{j}},x_{\hat{j}})\},\mathsf{fee})$. After receiving the $\mathsf{route}_{\mathsf{post}}$ message with $\{(\pi_{\hat{j}},x_{\hat{j}})\}$ from a entire candidate route. The recipient calls the function to send $\mathsf{Response}$ message $\langle\{(\pi_{\hat{j}},x_{\hat{j}})\},\mathsf{fee}\rangle$ to the sender.
    \item $\mathsf{path}_{\mathsf{opt}}$ $\leftarrow$ $\mathcal{S}.\mathsf{RangeRank}(\mathsf{totalFee})$. This algorithm selects the best route $\mathsf{path}_{\mathsf{opt}}$ from $\mathsf{totalFee}$ based on the forwarding fees.  In common practice, $\mathsf{RangeRank}$ is to choose the most economical route. Sometimes a sender might choose a route with a high forwarding fee, which can incentive intermediate nodes to prioritize his transactions. 
\end{itemize}

In a nutshell, zk-IPCN improves zk-PCN in two ways. zk-IPCN stores less channel information compared with zk-PCN. Moreover, zk-IPCN lowers the frequency of generating and broadcasting proofs. Thus the computing overhead and bandwidth usage of the network are reduced.
zk-IPCN also proves that zk-PCN can be compatible with other routing algorithms to improve the performance.

\section{Simulation Results}\label{impletation}

In this section, we investigate the performance of zk-PCN and zk-IPCN on the real Lightning Network topology. We compare the transaction success rates of zk-PCN and LN. Then we show the impact of the proof generation time and proof reachability on success rates. Finally, we show how zk-IPCN improves the performance of zk-PCN.

\subsection{Simulation Methods}\label{evaluation method}

The most crucial performance metric of the payment channel network is transaction success rates. Our experiments explore the impact of different factors on transaction success rates. 
We use the real-world Lightning Network (LN) topology to evaluate the performance of our scheme. The LN network \footnote{We get the snapshot of LN topology on 2021-03-31 (from \href{https://ln.fiatjaf.com/}{https://ln.fiatjaf.com/}).} contains 10529 nodes and 38910 channels.
The initial balances of each channel are invisible in LN, so we set up the initial balance as half of the channel capacity in our simulations. To implement zk-SNARKs, we use Aleo \footnote{\href{https://github.com/AleoHQ/aleo}{https://github.com/AleoHQ/aleo}} for circuit generation, proof generation and verification. Aleo uses the Groth16 algorithm \cite{groth2016size} on the backend to generate and verify proofs. 

Our simulation considers different channel capacity factors, defined as the times that each channel capacity is expanded. This is to explore how capacity impacts success rates. Besides, our simulation considers two types of transaction flows, uniform and skewed. In the uniform case, sender-recipient pairs are randomly and uniformly selected in the network. Skewed cases simulate the scenarios when the sender-recipient pairs are not uniformly distributed. Instead, the probability of nodes being chosen as a sender follows the exponential distribution $(\mathsf{skewness}/\mathsf{N})*e^{-x(\mathsf{skewness}/\mathsf{N})}$, where $N$ is the network size, and $\mathsf{skewness}$ is the skewness factor. The higher the skewness factor, the greater the probability that a small fraction of channels is utilized.
We simulate transaction amounts by randomly sampling them from LN \footnote{We sample the Bitcoin trace from 2021-03-01 to 2021-03-31.}. Since LN mainly proceeds micro transactions with high frequency and low transaction amounts, we simulate transaction amounts randomly in (0, the median of channel capacity).

In our experiment, we test the performance of zk-PCN, zk-IPCN and LN. Our experimental environment is equipped with Intel Xeon(R) Silver 4214R CPU and 98.2 GB RAM running 64-bit Ubuntu 20.04.3. We simulate ten times and present the average for each data point. The number of transactions in each simulation if not specified is 50000. All scatter points come with error bars. Some error bars are undersized and thereby not displayed.
\subsection{zk-PCN vs. LN}\label{LN}

We simulate zk-PCN on LN topology and measure transaction success rates of zk-PCN when the channel capacity factor or the payment skewness factor varies.

\begin{figure}[!htbp]
\centering
\subfigure[]{
\includegraphics[width=0.48\columnwidth]{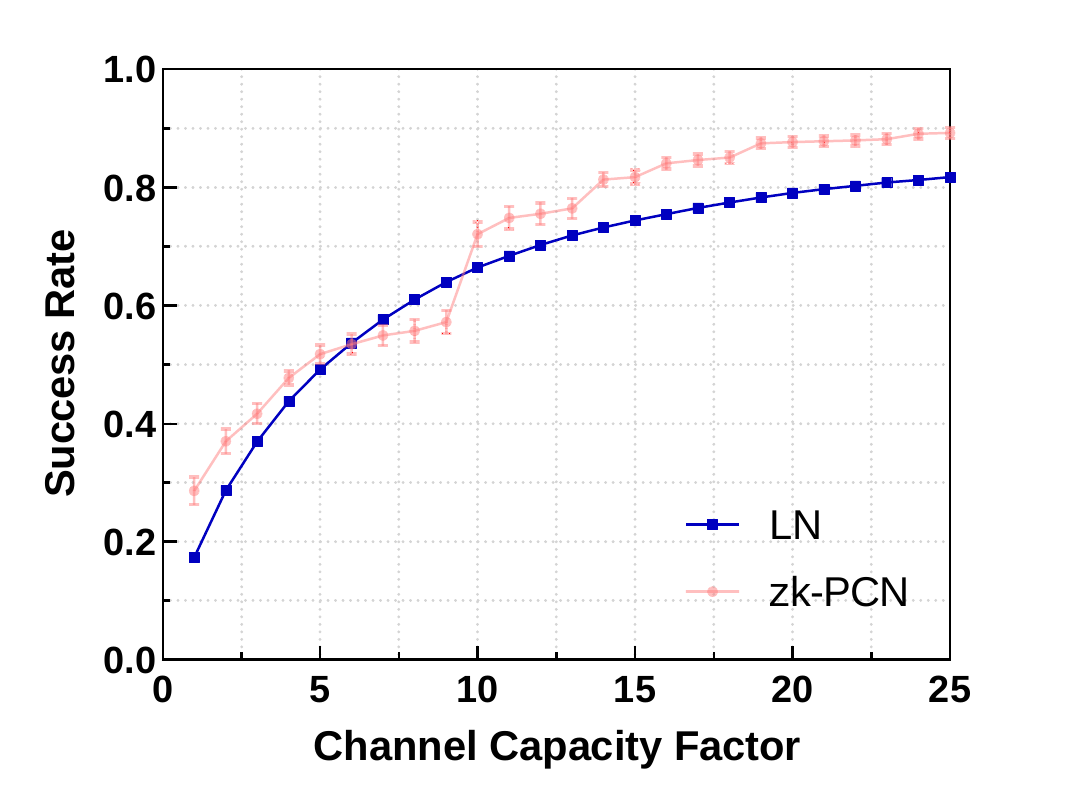}
\label{zk-PCN uniform performance}
}%
\subfigure[]{
\includegraphics[width=0.48\columnwidth]{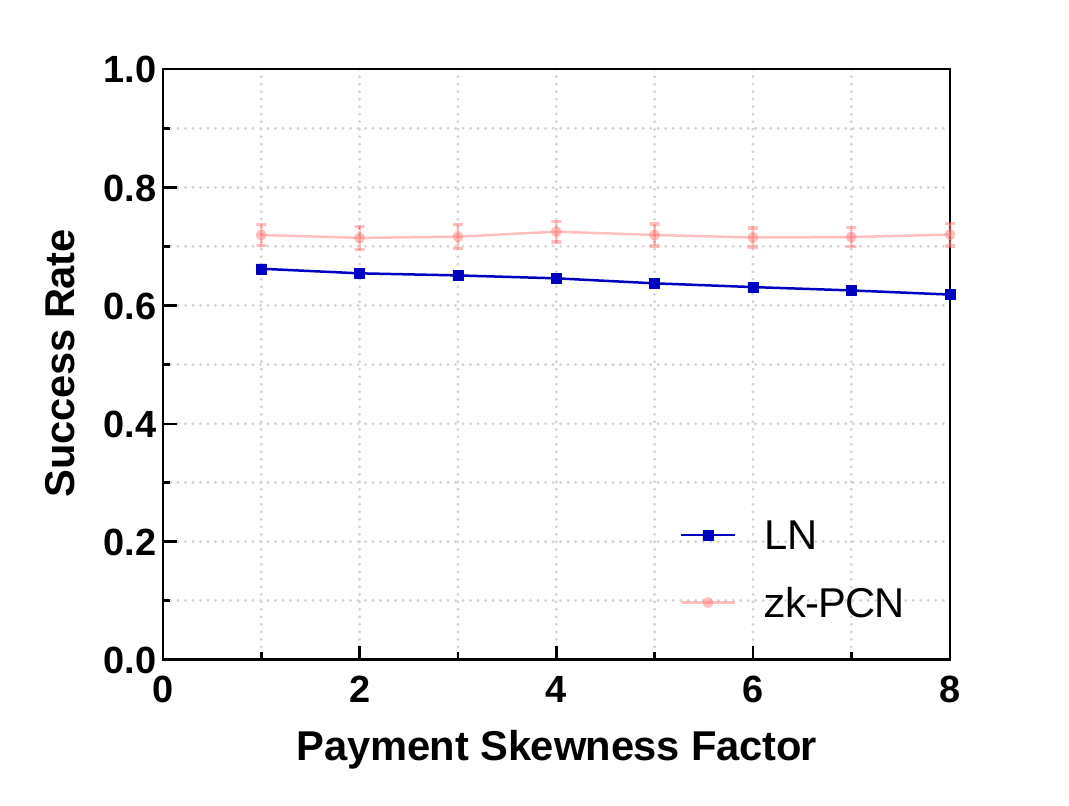}
\label{zk-PCN skewness performance}
}%
\centering
\caption{Success rates of LN and zk-PCN with varying (a) channel capacity and (b) payment skewness. }
\end{figure}

\textbf{Success rate with channel capacity.}  We vary the channel capacity from 1 to 25 and show the success rate of uniform payments in Fig.~\ref{zk-PCN uniform performance}. The success rate of zk-PCN is higher than that of LN in most cases. In the best-case scenario, zk-PCN outperforms LN by 11\%. The results indicate that opening public balance helps improve the success rate. The reason for failures in zk-PCN is that a sender cannot find an available route due to some capacity limitations. When the channel capacity factor increases, the success rate of zk-PCN can approach to 100\%.  

\textbf{Success rate with payment skewness.} We test zk-PCN's performance with skewed payments when the channel capacity factor is 10. We vary the skewness factor from 1 to 8 and show the success rate of skewed payments in Fig.~\ref{zk-PCN skewness performance}. The success rate of LN slightly decreases. However, the success rate of zk-PCN is not impacted by the payment skewness factor, and is higher than LN's success rate by 5\% - 10\%. Since zk-PCN selects the transaction route based on public balances, the success rate is not obviously affected by the payment skewness factor. LN prefers to choose the shortest path to forward the transaction. With the skewness factor increasing, short paths are used heavily and some balances on the path are quickly exhausted. Therefore, LN is easier to be affected by the skewness factor.

\subsection{The Performance of zk-PCN}\label{zk-PCN performance}

zk-PCN utilizes zk-SNARKs to generate proofs, so we explore the impact of zero-knowledge proofs on transaction success rates. We narrow the scope of LN topology and simulate the zk-PCN on 1000 nodes. 

\begin{table}[h]
\centering
\caption{Performance of generating and verifying proofs in zk-PCN}\label{overhead of proof}
\resizebox{\columnwidth}{3em}{
\begin{tabular}{|c|c|c|c|c|}
\hline
\textbf{Hash Times} & 1 & 10 & 100 & 1000 \\\hline
\textbf{Prover Time} & 157 ms & 682 ms & 6011 ms & 43798 ms\\\hline
\textbf{Verifier Time} & 5 ms & 5 ms & 6 ms & 5 ms\\
\hline
\textbf{Proof Size} & 193 B & 193 B & 193 B & 193 B\\
\hline
\end{tabular}
}
\end{table}

\begin{figure*}[!htbp]
\centering
\subfigure[Uniform]{
\includegraphics[width=0.24\textwidth]{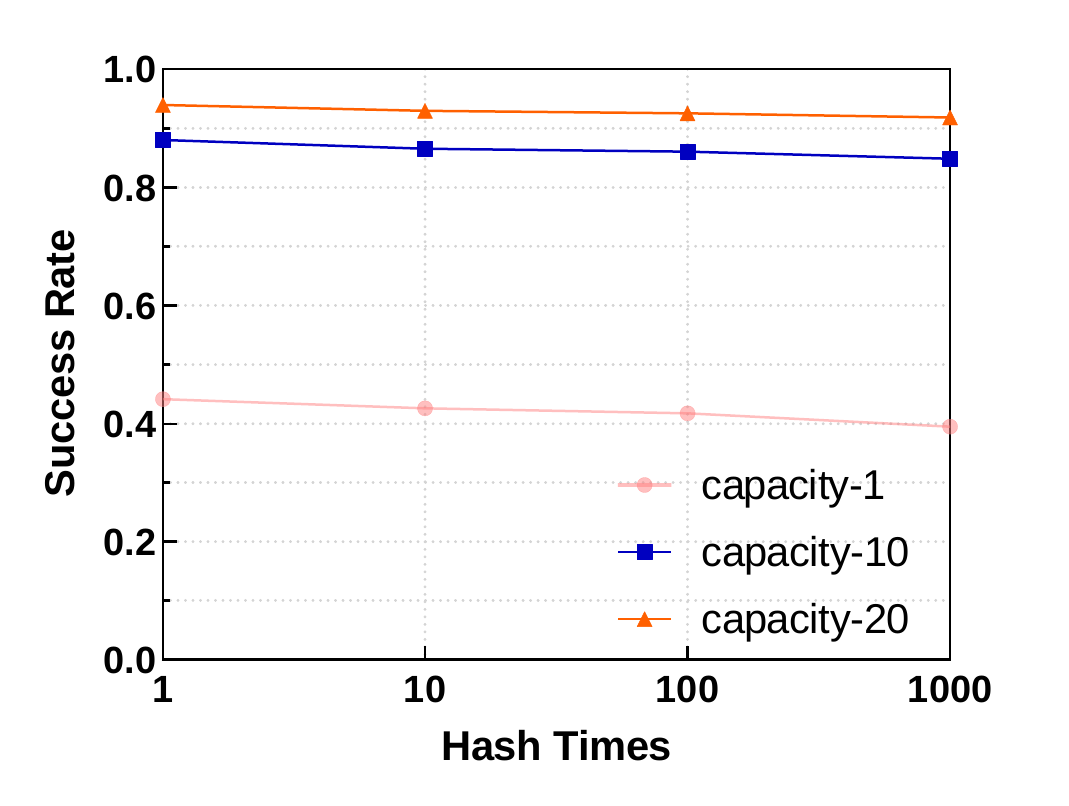}
\label{zk-PCN publicbal under capacity}
}%
\subfigure[Skewed]{
\includegraphics[width=0.24\textwidth]{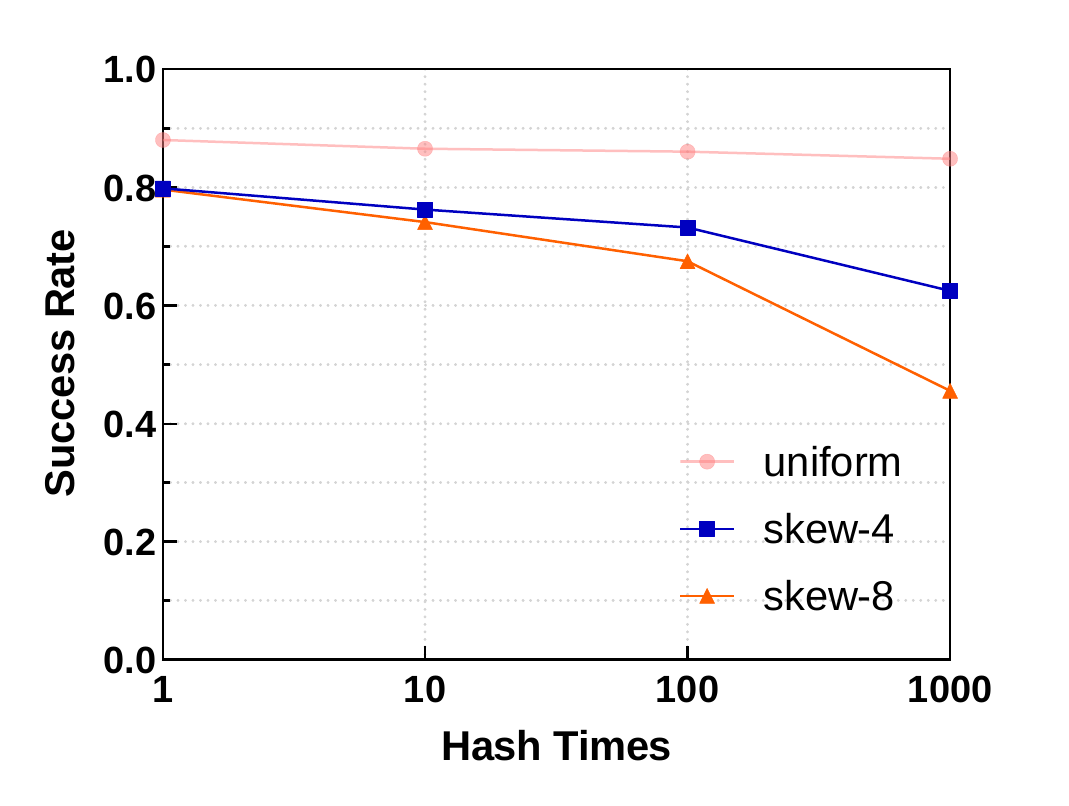}
\label{zk-PCN publicbal under skew}
}%
\subfigure[Uniform]{
\includegraphics[width=0.24\textwidth]{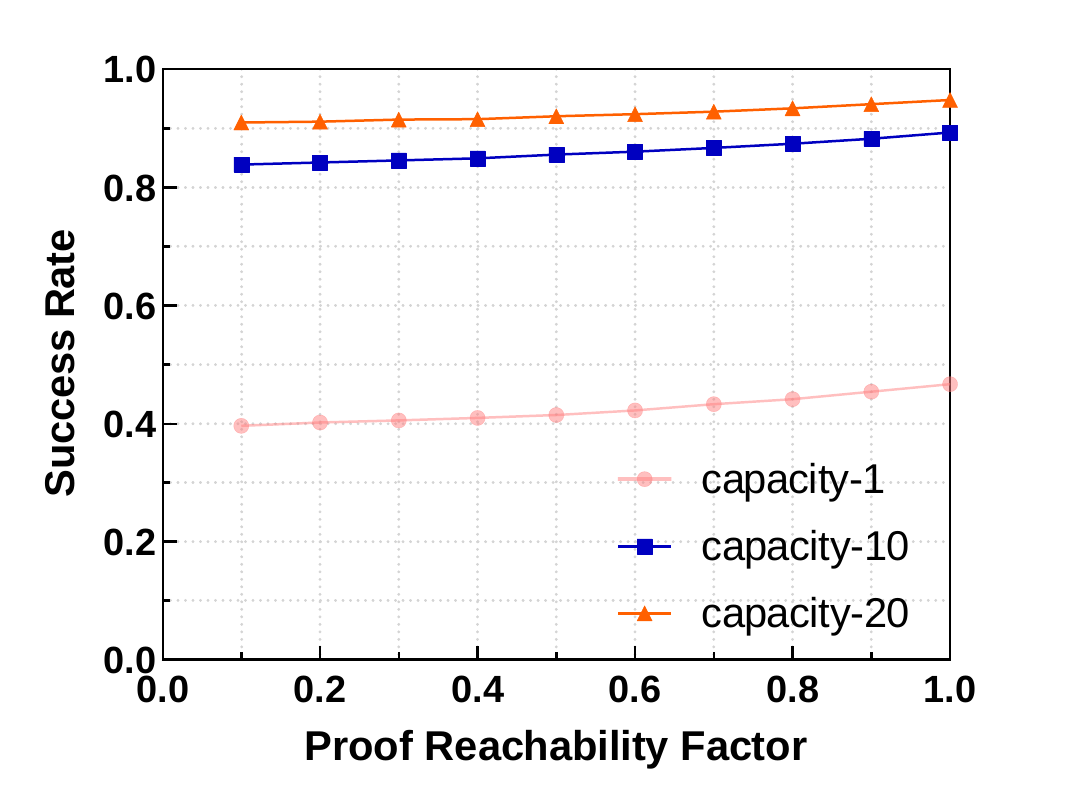}
\label{zk-PCN reach. under capacity}
}%
\subfigure[Skewed]{
\includegraphics[width=0.24\textwidth]{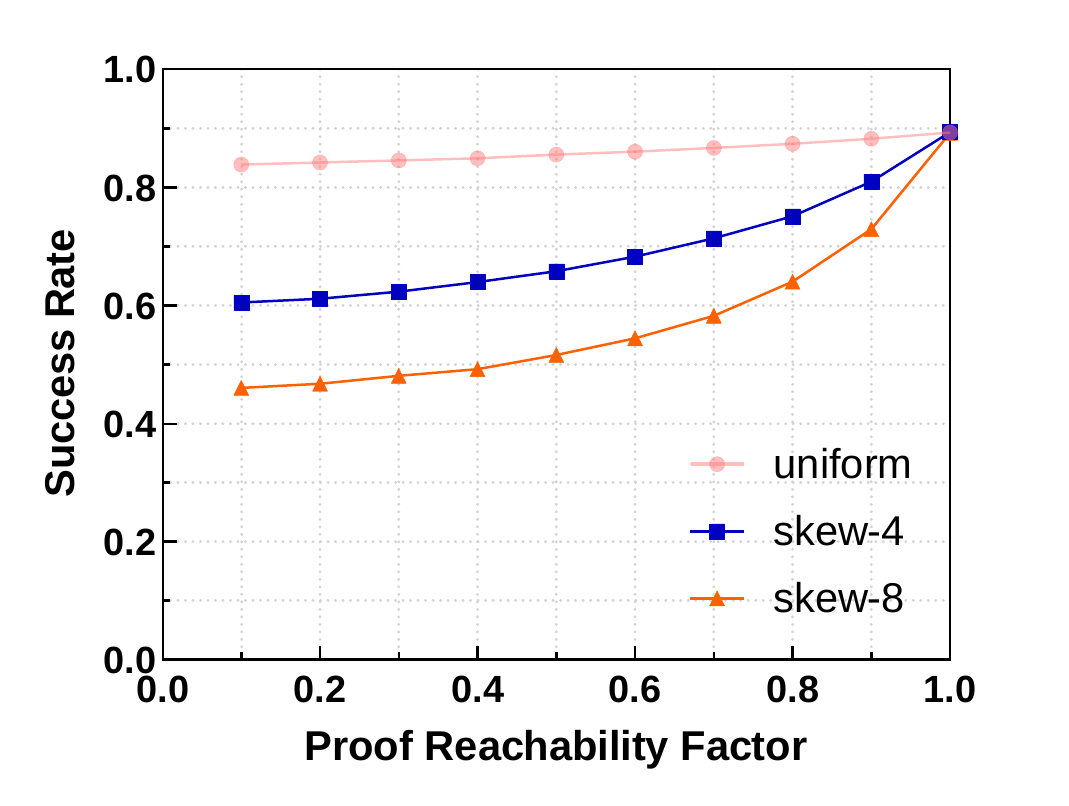}
\label{zk-PCN reach. under skew}
}%
\centering
\caption{Performance of zk-PCN with the varying number of hash times and proof reachability factor. }
\end{figure*}

\textbf{Performance of proof generation and verification.}
We vary the number of hash times (a hash represents a confirmed transaction) and show the prover time (proof generation), verifier time (proof verification) and proof size in TABLE~\ref{overhead of proof}. 
With the increase of hash times, the prover time linearly grows. We need 43798 ms to generate a proof for 1000 transactions. The verifier time is only on the order of milliseconds. The proof size is only of 193 bytes. When the number of transactions in the channel exceeds 1000, the time cost of generating a proof may be unacceptable. In such circumstance, nodes in a channel can reset the initial balances. 

\textbf{Success rate with proof generate time.} We test the success rate of zk-PCN performance with varying proof generation time measured by hash times. 
During proof generation, the public balances of a channel might not be up to date. When a sender selects this channel in an optimal route based on the old public balance, the transaction might fail because of insufficient funds. 
We range hash times from 1 to 1000. Fig.~\ref{zk-PCN publicbal under capacity} shows that when channel capacity is properly chosen, the success rate only reduces by 2\% - 5\%. In uniform cases, it is highly possible that channels are used bi-directionally so the range of balance variation is small. Even if it takes time to generate and broadcast proofs, balances can still afford incoming transactions. 
In Fig.~\ref{zk-PCN publicbal under skew}, we present the results under three cases including a uniform case, and two skewed cases (skewness is 4 and 8). When skewness is higher, the success rate is impacted more severely. In skewed cases, channels are mostly used unidirectionally. Some channels become unavailable due to insufficient true balances, but senders might still choose those channels whose public balances are not up to date due to long proof generation latency.

\textbf{Success rate with proof reachability.} 
We test zk-PCN with varying proof reachability factors. The reachability factor is the probability that a node can successfully receive a proof. This factor is to simulate network connection failures and faulty behaviours. If losing proofs, nodes on zk-PCN fail to update their routing tables, which could result in a decline in success rate. Fig.~\ref{zk-PCN reach. under capacity} shows that the success rate only decreases by at most 0.04 with the reachability factor ranging from 1.0 to 0 when capacity is 20. In Fig.~\ref{zk-PCN reach. under skew}, when skewness is 8, the success rate decreases by at most 0.43. When the proof reachability is 1, the success rate is not affected by the skewness.

\subsection{The Performance of zk-IPCN}\label{zk-IPCN performance}

\begin{figure}[!htbp]
\centering
\subfigure[Uniform]{
    \includegraphics[width=0.46\columnwidth]{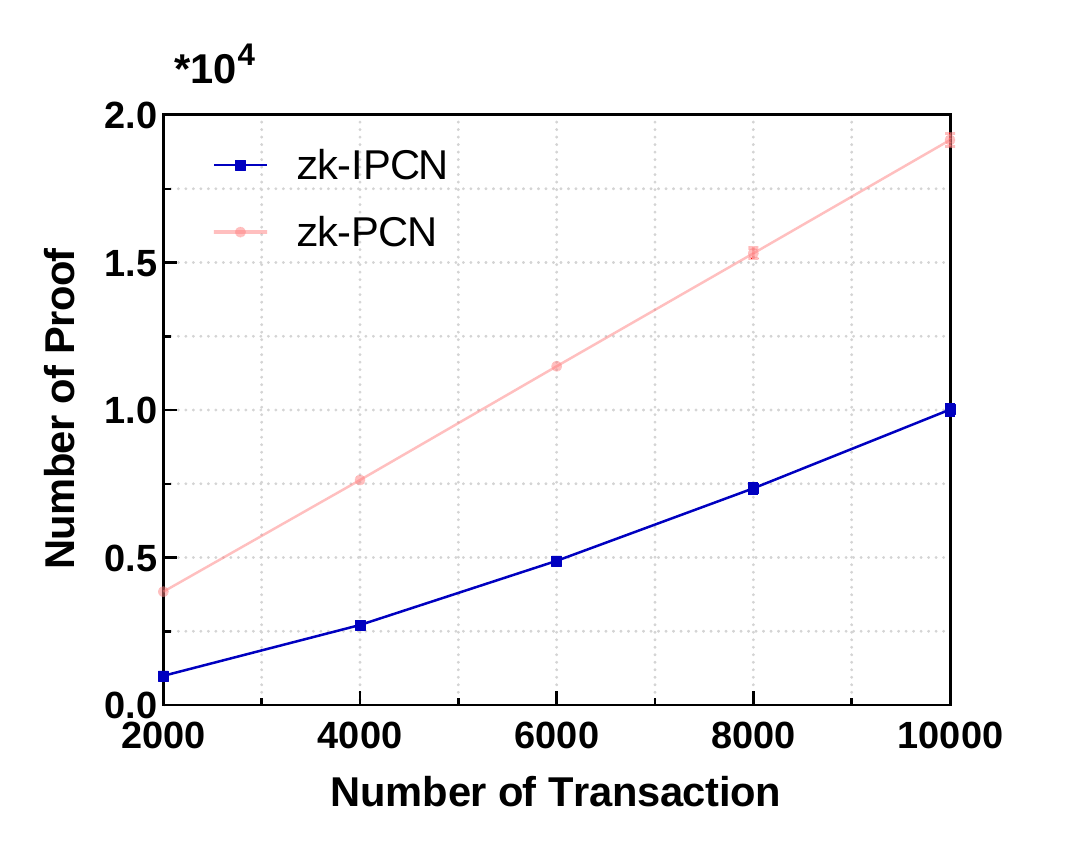}
    \label{zk-IPCN improved performance(uniform).}
}
\subfigure[Skewed]{
    \includegraphics[width=0.46\columnwidth]{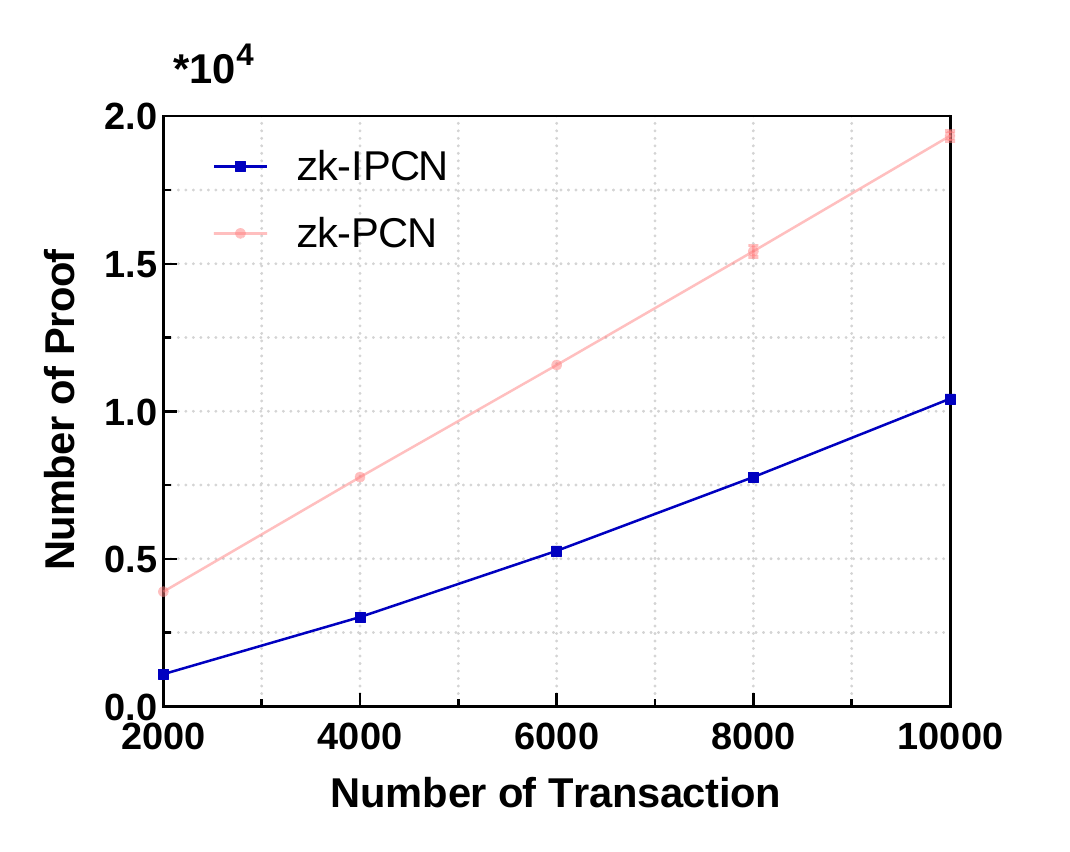}
    \label{zk-IPCN improved performance(skewed).}
}%
\caption{The number of proofs generated by zk-IPCN and zk-PCN in (a) uniform (b) skewed cases.}
\end{figure}

We simulate zk-IPCN and measure the number of proof generated by zk-IPCN and zk-PCN when the transaction number changes.
We test the number of proof generated by zk-IPCN and zk-PCN with varying the number of transactions. In the uniform case as shown in Fig.~\ref{zk-IPCN improved performance(uniform).}, the slopes of zk-IPCN and zk-PCN are 1.127 and 1.914, respectively. The nodes in zk-IPCN generate proof only after receiving the $\mathsf{route}_{\mathsf{post}}$ message, without having to generate and broadcast proofs for each transaction. In Fig~\ref{zk-IPCN improved performance(skewed).}, the slopes of zk-IPCN and zk-PCN are 1.166 and 1.931. Compared with the uniform case, the number of proofs generated increases by at most 400 under the skewed case. After controlling the transaction flow, the length of the transaction path increases, resulting in an increase in the number of proofs generated.

\section{Conclusion}\label{conslusion}

In this paper, we propose zk-PCN, a privacy-preserving payment channel network that can protect balance secrecy, relationship anonymity and payment privacy properties. It can also resist the balance discovery attack and the payment discovery attack. zk-PCN introduces public balances to hide true balances with an authenticity guarantee based on zk-SNARKs. zk-PCN achieves a high transaction success rate compared with LN. We also propose zk-IPCN, an improved version of zk-PCN. zk-IPCN adopts a proof generation algorithm (RPG) to reduce the frequency of proof generation and nodes' local storage overheads. Finally, simulations demonstrate that zk-PCN achieves high transaction success rates under various settings, and zk-IPCN improves the performance of zk-PCN.

\section{Acknowlegement}\label{ack}

This study was paritially supported by the National Natural Science Foundation of China under grants 61832012, 62102232 and 62122042, the National Research Foundation, Singapore and Infocomm Media Development Authority under its Future Communications Research \& Development Programme. 

\bibliographystyle{IEEEtran}
\bibliography{IEEEabrv, ref}

\end{document}